# Computational Performance and Energy Efficiency
# of ARM based HPC servers

Oskar Schirmer

2023-07-07

## Abstract

HPC world is dominated by x86 ISA CPUs. This monoculture is not necessarily justified by best performance evaluation, but may inherit from e.g. SW related restrictions on the choice of HW platforms. To avoid running (further) into path dependency, alternate HW platforms need to be evaluated for performance compared to existing HPC setup. As a result, it may turn out alternate HW platforms are more efficient for HPC. In any case, even if performance differences are low, avoiding path dependencies that stem from HW choice restrictions simplifies switching to different HW platforms in future, should suitable systems evolve. Moreover, broadening the perspective to generic HW platforms may trigger cooperation and wield influence on HW platform development, resulting in HW/SW co-design advantages.

# Contents

# 1 Hardware

Two different ARM based HPC servers to be investigated are:

- NVidia Arm HPC Developer Kit:[1]
  model: GIGABYTE G242-P32, 2U server
  CPU: 1x Ampere Altra Q80-30, 80 cores/CPU
  memory: 512G DDR4
  GPU: 2x NVidia A100
  DPU: 2x NVidia BlueField-2 E-Series

- HAICGU cluster at FIAS (Universität Frankfurt am Main):[2]
  model: TaiShan 200 (Model 2280)
  CPU: 2x ARM Kunpeng 920, 64 cores/CPU, 2.6 GHz
  memory: 128G DDR4

  AI Training Node, model: Atlas 800 (Model 9000)
  CPU: 4x ARM Kunpeng 920
  memory: 1024G (32x 32GB DDR4 2933MHz RDIMM)
  NPU: 8x Huawei Ascend 910 with 32 AI cores, 32GB HBM2 memory

  AI Inference Node, model: Atlas 800 (Model 3000)
  CPU: 2x ARM Kunpeng 920
  memory; 512G (16x 32GB DDR4 2933 MHz RDIMM)
  GPU: 5x Atlas 300 AI Inference Card; 32GB; PCIe3.0 x16

These are compared against production systems at GWDG:

- SCC (partition medium)
  model: Cascade Lake
  CPU: 2x Xeon Platinum 9242, 48 cores/CPU, 3.8 GHz
  memory: 384G

- SCC/GPU (partition gpu)
  model: Broadwell
  CPU: 2x Xeon E5-2650 v4, 12 cores/CPU, 2.2 GHz
  memory: 128G
  GPU: 2x GTX 1080

---

[1] https://developer.nvidia.com/arm-hpc-devkit
[2] https://www.open-edge-hpc-initiative.org/undertakings/arm-systems

- HLRN Emmy Phase 1 (partition medium40)
  model:
  CPU: 2x Intel Xeon Gold 6148 SKL-SP, 20 cores/CPU
  memory: 187G
  GPU: 4x Tesla V100

- HLRN Emmy Phase 2 (partition standard96)
  CPU: 2x Xeon Platinum 9242 CLX-AP, 48 cores/CPU
  memory: 376G

- HLRN Grete (partition grete)
  model:
  CPU: 2x AMD EPYC 7513, 32 cores/CPU
  memory: 512G
  GPU: 4x Nvidia A100

# 2 Software

"GROMACS is a full-featured suite of programs to perform molecular dynamics simulations. GROMACS can run in parallel on multiple cores of a single workstation using its built-in thread-MPI.", see [2019kp].[3] With the standardized set of input data, "benchRIB",[4] it is used as a benchmark tool for multicore systems.

While it seems unpopular to document on how tests were setup and run in detail (see e.g. [2022kk], [2023td]), it is of course crucial to give all details that would allow replication of the results for verification, as is demanded at [2022df].

To retrieve and prepare the necessary source files:

```
wget --no-check-certificate https://ftp.gromacs.org/gromacs/gromacs-2022.5.tar.gz
tar -xzf gromacs-2022.5.tar.gz
wget --no-check-certificate https://ftp.gromacs.org/regressiontests/regressiontests-2022.5.tar.gz
tar -xzf regressiontests-2022.5.tar.gz
wget https://www.mpinat.mpg.de/benchRIB.zip
unzip benchRIB.zip
```

Configuration depends on the available hardware:

## 2.1 NVidia ARM

Operating system run on the NVidia ARM nodes: Rocky Linux 8.6.
Configuration without GPU support:

---

[3] https://manual.gromacs.org
[4] https://www.mpinat.mpg.de/benchRIB

```
mkdir -p ~/opt
mkdir -p gromacs-2022.5/build
cd gromacs-2022.5/build
cmake .. -DGMX_BUILD_OWN_FFTW=OFF -DCMAKE_C_COMPILER=mpicc -DCMAKE_CXX_COMPILER=mpic++ \
        -DREGRESSIONTEST_DOWNLOAD=OFF -DREGRESSIONTEST_PATH=../../regressiontests-2022.5/ \
        -DGMX_MPI=on -DCMAKE_INSTALL_PREFIX=${HOME}/opt
make -j
make install
```

For configuration with GPU support, instead:

```
CC=gcc CXX=c++ cmake .. -DGMX_BUILD_OWN_FFTW=OFF -DGMX_SIMD=AUTO -DGMX_MPI=OFF \
        -DREGRESSIONTEST_DOWNLOAD=OFF -DREGRESSIONTEST_PATH=../../regressiontests-2022.5/
        -DGMX_GPU=CUDA -DGMX_OPENMP=ON -DCMAKE_INSTALL_PREFIX=${HOME}/opt \
        -DGMX_INSTALL_NBLIB_API=OFF -DGMX_THREAD_MPI=ON \
        -DCUDA_TOOLKIT_ROOT_DIR=/opt/nvidia/hpc_sdk/Linux_aarch64/23.1/cuda \
        -DCMAKE_CXX_FLAGS="-I /opt/nvidia/hpc_sdk/Linux_aarch64/23.1/math_libs/include/" \
        -DCMAKE_C_FLAGS="-I /opt/nvidia/hpc_sdk/Linux_aarch64/23.1/math_libs/include/" \
        -DCMAKE_CXX_LINK_FLAGS="-Wl,-rpath,/opt/nvidia/hpc_sdk/Linux_aarch64/23.1/math_libs/lib64/" \
        -DCMAKE_C_LINK_FLAGS="-Wl,-rpath,/opt/nvidia/hpc_sdk/Linux_aarch64/23.1/math_libs/lib64/"
```

It may be necessary to provide existing libraries into the expected search path, e.g.:

```
sudo ln -s /opt/nvidia/hpc_sdk/Linux_aarch64/23.1/math_libs/lib64/libcufft.so.11 /opt/nvidia/hpc_sdk/Linux_aarch64/23.1/cuda/lib64/
sudo ln -s /opt/nvidia/hpc_sdk/Linux_aarch64/23.1/math_libs/lib64/libcufft.so /opt/nvidia/hpc_sdk/Linux_aarch64/23.1/cuda/lib64/
```

## 2.2 HAICGU

Operating system run on the HAICGU cluster: Rocky Linux 8.7.

Configuration without GPU support:

```
module load GCC/9.5.0
module load CMake/3.23.1
module load OpenMPI/4.1.3
module load FFTW/3.3.10
mkdir -p ~/opt
mkdir -p gromacs-2022.5/build
cd gromacs-2022.5/build
cmake .. -DGMX_BUILD_OWN_FFTW=OFF -DCMAKE_C_COMPILER=mpicc -DCMAKE_CXX_COMPILER=mpic++ \
        -DREGRESSIONTEST_DOWNLOAD=OFF -DREGRESSIONTEST_PATH=../../regressiontests-2022.5/ \
        -DGMX_MPI=on -DCMAKE_INSTALL_PREFIX=${HOME}/opt
make -j
make install
```

Currently there is no GPU support available for GROMACS on the HAICGU cluster.

## 2.3 SCC

Operating system run on the SCC cluster: Scientific Linux 7.9.

Configuration without GPU support:

```
module load cmake/3.21.4
module load gcc/9.3.0
module load openmpi/4.1.1
mkdir build && cd build
cmake .. -DGMX_MPI:BOOL=ON -DCMAKE_C_COMPILER=/opt/sw/spack/0.17.1/lib/spack/env/gcc/gcc \
    -DCMAKE_CXX_COMPILER=/opt/sw/spack/0.17.1/lib/spack/env/gcc/g++ \
    -DMPI_C_COMPILER=/opt/sw/rev/21.12/cascadelake/gcc-9.3.0/openmpi-4.1.1-ibhhql/bin/mpicc \
    -DMPI_CXX_COMPILER=/opt/sw/rev/21.12/cascadelake/gcc-9.3.0/openmpi-4.1.1-ibhhql/bin/mpic++ \
    -DGMX_INSTALL_LEGACY_API=ON -DGMX_HWLOC:BOOL=ON -DGMX_GPU:STRING=OFF \
    -DGMX_EXTERNAL_LAPACK:BOOL=OFF -DGMX_EXTERNAL_BLAS:BOOL=OFF -DGMX_SIMD:STRING=AVX_512 \
    -DGMX_USE_RDTSCP:BOOL=ON -DGMX_OPENMP:BOOL=ON -DGMX_CYCLE_SUBCOUNTERS:BOOL=OFF \
    -DGMX_FFT_LIBRARY=fftw3 -DGMX_VERSION_STRING_OF_FORK=spack
make -j
```

Configuration with GPU support:

```
module load gcc/9.3.0
module load cmake/3.21.4
module load python/3.9.0
module load cuda/11.5.1
module load fftw/3.3.10
mkdir -p ~/opt
mkdir -p gromacs-2022.5/build
cd gromacs-2022.5/build
CC=gcc CXX=c++ cmake .. -DGMX_BUILD_OWN_FFTW=OFF -DGMX_SIMD=AUTO -DGMX_MPI=OFF -DGMX_GPU=CUDA -DGMX_OPENMP=ON \
        -DREGRESSIONTEST_DOWNLOAD=OFF -DREGRESSIONTEST_PATH=../../regressiontests-2022.5/ \
        -DCMAKE_INSTALL_PREFIX=${HOME}/opt -DGMX_INSTALL_NBLIB_API=OFF -DGMX_THREAD_MPI=ON
make -j
make install
```

## 2.4 HLRN

Operating system run on the HLRN nodes without GPU: CentOS Linux 7, operating system run on the HLRN nodes with GPU: Rocky Linux 8.7.

Configuration without GPU support:

```
module load cmake/3.26.4
module load gcc/9.3.0
module load intel/2022.2
[[ ! -f cmake/FindMKL.cmake ]] && \
    wget https://raw.githubusercontent.com/pytorch/pytorch/main/cmake/Modules/FindMKL.cmake -O cmake/FindMKL.cmake
mkdir build && cd build
source /sw/tools/oneapi/2022.2/compiler/latest/env/vars.sh intel64
source /sw/tools/oneapi/2022.2/mpi/latest/env/vars.sh intel64
cmake .. -DGMX_FFT_LIBRARY=mkl -DCMAKE_C_COMPILER=mpiicc -DCMAKE_CXX_COMPILER=mpiicpc \
    -DCMAKE_INSTALL_PREFIX=/sw/chem/gromacs/2022.5/skl/impi -DGMX_GPU=off -DGMX_MPI=on \
    -DGMX_BUILD_SHARED_EXE=OFF -DMKL_INCLUDE_DIR=/sw/tools/oneapi/2022.2/mkl/latest/include \
    -DCMAKE_CXX_LINK_FLAGS="-Wl,-rpath,/sw/tools/oneapi/2022.2/compiler/latest/linux/compiler/lib/intel64_lin,\
-rpath,/sw/tools/oneapi/2022.2/mkl/latest/lib/intel64,-rpath,/sw/compiler/gcc/9.3.0/skl/lib64" \
    -DCMAKE_C_LINK_FLAGS="-Wl,-rpath,/sw/tools/oneapi/2022.2/compiler/latest/linux/compiler/lib/intel64_lin,\
-rpath,/sw/tools/oneapi/2022.2/mkl/latest/lib/intel64" \
    -DGMX_SIMD=AVX_512 -DGMX_GPLUSPLUS_PATH=/sw/compiler/gcc/9.3.0/skl/bin/g++ -DGMXAPI=OFF \
    -DMKL_LIBRARIES=/sw/tools/oneapi/2022.2/mkl/latest/lib/intel64/libmkl_rt.so
make -j
```

Configuration with GPU support:

```
module load gcc/9.3.0
module load cmake/3.26.4
module load python/3.9.16
module load cuda/12.0
module load fftw3/3.3.8
mkdir -p ~/opt
mkdir -p gromacs-2022.5/build
cd gromacs-2022.5/build
CC=gcc CXX=c++ cmake .. -DGMX_BUILD_OWN_FFTW=OFF -DGMX_SIMD=AUTO -DGMX_MPI=OFF \
        -DREGRESSIONTEST_DOWNLOAD=OFF -DREGRESSIONTEST_PATH=../../regressiontests-2022.5/ \
        -DGMX_GPU=CUDA -DGMX_OPENMP=ON -DCMAKE_INSTALL_PREFIX=${HOME}/opt \
        -DGMX_INSTALL_NBLIB_API=OFF -DGMX_THREAD_MPI=ON
make -j
make install
```

# 3 Benchmark Invocation

To run GROMACS benchmark with no GPU support, the environment is setup as follows, run on a single node, using all cores available. Invocation of the test itself includes measuring the time to complete. *cores-per-node* is choosen to match the hardware, using the maximum number of cores available per node, without deploying hyperthreading:

| system | CPU/node | cores/CPU | *cores-per-node* | hyperthreading |
|---|---|---|---|---|
| NVidia ARM | 1 | 80 | 80 | none |
| HAICGU | 2 | 64 | 128 | none |
| SCC, medium | 2 | 48 | 96 | not enabled |
| SCC, gpu | 2 | 12 | 24 | not enabled |
| HLRN, medium40 | 2 | 20 | 40 | 2 |
| HLRN, standard96 | 2 | 48 | 96 | 2 |
| HLRN, grete | 2 | 32 | 64 | 2 |

## 3.1 NVidia ARM

Invocation without GPU support:

```
export SLURM_CPU_BIND=none
export OMP_NUM_THREADS=1
source ${HOME}/opt/bin/GMXRC
time mpirun -np 80 gmx_mpi mdrun -ntomp 1 -s benchRIB.tpr -nsteps 10000 -dlb yes -v
```

Invocation with GPU support:

```
export GMX_GPU_PME_PP_COMMS=true
export GMX_GPU_DD_COMMS=true
source ${HOME}/opt/bin/GMXRC
time gmx mdrun -nb gpu -pme gpu -npme 1 -bonded gpu -ntmpi 8 -ntomp 0 -s benchRIB.tpr \
     -pin on -gputasks 00001111 -nsteps 10000 -v
```

## 3.2 HAICGU

Invocation without GPU support:

```
module load OpenMPI/4.1.3
module load FFTW/3.3.10
export SLURM_CPU_BIND=none
export OMP_NUM_THREADS=1
source ${HOME}/opt/bin/GMXRC
time mpirun -np 128 gmx_mpi mdrun -ntomp 1 -s benchRIB.tpr -nsteps 10000 -dlb yes -v
```

## 3.3 SCC

Invocation on partition `medium`:

```
module load gromacs/2021.3
export SLURM_CPU_BIND=none
export OMP_NUM_THREADS=1
source $(which GMXRC)
time mpirun -np 96 gmx_mpi mdrun -ntomp 1 -s benchRIB.tpr -nsteps 10000 -dlb yes -v
```

Invocation on partition `gpu`:

```
module load cuda/11.5.1
module load fftw/3.3.10
unset OMP_NUM_THREADS
export GMX_GPU_PME_PP_COMMS=true
export GMX_GPU_DD_COMMS=true
export SLURM_CPU_BIND=none
source ${HOME}/opt/bin/GMXRC
time gmx mdrun -nb gpu -pme gpu -npme 1 -bonded gpu -ntmpi 8 -ntomp 0 -s scratch/benchRIB.tpr -pin on \
     -gputasks 00112233 -nsteps 10000 -v
```

## 3.4 HLRN

Invocation on partition `medium40` or `standard96`:

```
    module load impi/2021.6 gromacs/2021.2
export SLURM_CPU_BIND=none
export OMP_NUM_THREADS=1
source $(which GMXRC)
time mpirun -np cores-per-node gmx_mpi mdrun -ntomp 1 -s benchRIB.tpr \
     -nsteps 10000 -dlb yes -v
```

Invocation with GPU support on the `grete` partition:

```
module load gcc/9.3.0 cmake/3.26.4 python/3.9.16 cuda/12.0 fftw3/3.3.8
unset OMP_NUM_THREADS
export GMX_GPU_PME_PP_COMMS=true
export GMX_GPU_DD_COMMS=true
export SLURM_CPU_BIND=none
source ${HOME}/opt/bin/GMXRC
time gmx mdrun -nb gpu -pme gpu -npme 1 -bonded gpu -ntmpi 16 -ntomp 0 -s scratch/benchRIB.tpr \
     -pin on -gputasks 0000111122223333 -nsteps 10000 -v
```

# 4 Performance Evaluation

## 4.1 Run Time Measurement

- `gmx_mpi` produces time stamps for start and finish of computation. These do not include time to setup the processes (net).

- `time`, in contrast, measures the time to run `gmx_mpi`, including any process setup times (gros).

- Furthermore, `gmx_mpi` produces the *ns/day* value as a net computational performance indicator.

Net time measurement of computational intensive parts of the benchmark tool are useful only to compare computational performance of equivalent runs. When it comes to determining the energy consumption of a run, it is necessary to match the precise span of time to the power consumed meanwhile. For this reason, only the gros times are further evaluated.

## 4.2 Energy Consumption Measurement

Different tools are provided for energy measurement:

- NVidia ARM: `/usr/sbin/ipmi-dcmi --get-system-power-statistics`

- HAICGU: *not available*

- SCC: `/usr/sbin/ipmi-oem intelnm get-node-manager-statistics mode=globalpower`

- HLRN: `sacct -o ConsumedEnergyraw`

The `ipmi` tools show current power consumption, while `sacct` reports the total energy consumed for a task. Invocation of the `ipmi` tools during test run may affect the benchmark result, as it provokes context switches.

## 4.3 Heuristic Runtime Configuration

GROMACS uses a heuristic approach to find an optimized distribution of computational tasks to the available hardware. In doing so, it renders the benchmark result non-deterministic, and performance of running it with command line parameters unchanged on the same hardware may well differ by more than 15E.g., three subsequent runs on SCC yield the following configuration and results:

```
> grep '\(Performance:\|optimal\)' slurm-1598*
slurm-15980750.out:             optimal pme grid 208 208 208, coulomb cutoff 1.113
slurm-15980750.out:Performance:        8.776          2.735
slurm-15986805.out:             optimal pme grid 224 224 224, coulomb cutoff 1.033
slurm-15986805.out:Performance:       10.134          2.368
slurm-15987865.out:             optimal pme grid 240 240 240, coulomb cutoff 1.000
slurm-15987865.out:Performance:        9.797          2.450
```

Possible causes for the observed variations include:

- - changes in system temperature (including caused by ambient temperatures)

- - non-deterministic daemons and system services behaviour

- - non-deterministic data transfer behaviour (may well depend on behaviour of compute jobs on other nodes)

- - initial system state depending on previous job run on the node (i.e. non-deterministic node history)

- - component variation in node production (only relevant for different results from one node to another)

Taking these observations into account, it is clear that comparision of benchmarks from different systems needs at least to be treated with special caution.

## 4.4 Benchmark Run Duration

The value of nsteps 10000 has been chosen as a compromise to average over a sufficient amount of time on the one side. yet to achieve results in reasonable, and to avoid runtime failure, which occasionally occurs with nsteps larger than 100000.

In the latter case output is as follows:

```
[...]
Step 110300  Warning: pressure scaling more than 1%, mu: 1.01555 1.01555 1.01555
step 110300, remaining wall clock time:    106 s
Step 110301, time 441.204 (ps)   LINCS WARNING
relative constraint deviation after LINCS:
rms 61.759613, max 11656.978516 (between atoms 79208 and 79210)
bonds that rotated more than 30 degrees:
 atom 1 atom 2  angle  previous, current, constraint length

Step 110301, time 441.204 (ps)   LINCS WARNING
relative constraint deviation after LINCS:
rms 109.216400, max 12557.813477 (between atoms 79211 and 79210)
bonds that rotated more than 30 degrees:
 atom 1 atom 2  angle  previous, current, constraint length
```

```
 152394 152393   36.9   0.1379  0.1573    0.1358
 152370 152369   42.0   0.0975  0.1151    0.0960
[...]
 152350 152348   60.7   0.1432  0.2286    0.1410
 152356 152354   36.7   0.1371  0.1524    0.1350

step 110301: One or more water molecules can not be settled.
Check for bad contacts and/or reduce the timestep if appropriate.

step 110301: One or more water molecules can not be settled.
Check for bad contacts and/or reduce the timestep if appropriate.
Wrote pdb files with previous and current coordinates
Wrote pdb files with previous and current coordinates

WARNING: Could not unregister pinned host memory used for GPU transfers.  An unhandled error
         from a previous CUDA operation was detected. CUDA error #700 (cudaErrorIllegalAddress):
         an illegal memory access was encountered.
[...]
Program:     gmx mdrun, version 2022.5
Source file: src/gromacs/gpu_utils/pinning.cu (line 122)
Function:    gmx::unpinBuffer(void*)::<lambda()>
MPI rank:    0 (out of 8)

Assertion failed:
Condition: stat == cudaSuccess
Could not unregister pinned host memory used for GPU transfers. CUDA error
#700 (cudaErrorIllegalAddress): an illegal memory access was encountered.

For more information and tips for troubleshooting, please check the GROMACS
website at http://www.gromacs.org/Documentation/Errors
[...]
```

## 4.5 Performance Counter Handling

The option -resethway has not been used throughout the benchmarking, because it would occasionally trigger errors, producing output as follows:

```
[...]
Program:     gmx mdrun, version 2022.5
Source file: src/gromacs/mdlib/resethandler.cpp (line 161)
MPI rank:    2 (out of 16)

Fatal error:
PME tuning was still active when attempting to reset mdrun counters at step
5000. Try resetting counters later in the run, e.g. with gmx mdrun -resetstep.

For more information and tips for troubleshooting, please check the GROMACS
website at http://www.gromacs.org/Documentation/Errors
[...]
```

Another reason not to use the `-resethway` option is it reduces the performance measurement to only part of the benchmark run, which does not match the span of time for which energy consumption is evaluated, Doing so would invalidate any energy efficiency conclusions.

# 5 Results

The following results are preliminary in that calculation of run time is done differently on different systems, thus not directly comparable. Each test run on a single node.

| system | target | gromacs | time[s] | ns/day | power[W] | energy[kJ]/ns |
|---|---|---|---|---|---|---|
| NVidia ARM | `nv-arm02` | 2022.5 | 774 | 4.631 | 595.8 | 11116 |
| NVidia ARM/GPU | `nv-arm02` | 2022.5/cuda | 141 | 29.366 | 710.4 | 2091 |
| NVidia ARM/DPU | | not supported | | | | |
| HAICGU | `cn07` | 2022.5 | 981 | 3.703 | not available | |
| HAICGU CANN | | not supported | | | | |
| SCC | `amp060` | 2021.3 | 395 | 10.134 | 887.7 | 7569 |
| SCC/gtx1080 | `dge003` | 2022.5/cuda | 364 | 10.454 | not available | |
| HLRN medium40 | `gcn1347` | 2022.5 | 774 | 4.579 | 525.8 | 9921 |
| HLRN standard96 | `gcn2335` | 2022.5 | 358 | 10.177 | 837.3 | 7109 |
| HLRN grete/GPU | `ggpu105` | 2022.5/cuda | 143 | 29.500 | 773.5 | 2266 |

# 6 Conclusion

The GROMACS software is bound to specific libraries for support of hardware acceleration such as GPU and similar. Different libraries are implemented for different hardware, not all of these are prepared for use on the GROMACS implementation side. For this reason, both NVidia ARM DPU and HAICGU AI accelerator are not available with GROMACS. It would need major software implementation effort to provide appropriate support.

Performance measurements include variation of some 15% from one test run to the next identical test (i.e. same configuration on same hardware). Consequently, it does not make sense to look at numbers as precise, but only as an indication of the order of magnitude of some performance. To obtain more precise results, it would need extensive test run series to lower the corresponding thresholds of significance, see [2006bd].

The results of measurement for test runs without GPU suggest that computational performance directly correlates to the number of cores per node, with ARM based systems yielding only about a quarter of the performance per core.

For test runs including GPU, grete nodes yield about three times the performance of what we see on SCC, but the Nvidia ARM with only two GPUs shows about the same performance as do the grete nodes with four GPUs installed. It is not immediately clear why this is, more so as both systems feature A100 GPUs.

Concerning energy consumption, using GPUs reduces power consumption by an order of magnitude. As one would expect, power consumption for the Nvidia ARM based system is much the same as for grete, as both systems deploy A100 GPUs.